Title: **Simulations Suggest Pharmacological Methods for Rescuing Long-Term Potentiation**

Authors: Paul Smolen, Douglas A. Baxter, and John H. Byrne

21 pages, 4 figures

Laboratory of Origin:
Department of Neurobiology and Anatomy
W. M. Keck Center for the Neurobiology of Learning and Memory
The University of Texas Medical School at Houston
Houston, Texas 77030
United States of America

Running Title: Pharmacological Approaches to Rescue LTP

Correspondence Address:
Paul Smolen
Department of Neurobiology and Anatomy
The University of Texas-Houston Medical School
Houston, TX 77030
E-mail: Paul.D.Smolen@uth.tmc.edu
Voice: (713) 500-5564
FAX: (713) 500-0621

## ABSTRACT

Congenital cognitive dysfunctions are frequently due to deficits in molecular pathways that underlie the induction or maintenance of synaptic plasticity. For example, Rubinstein-Taybi syndrome (RTS) is due to a mutation in *cbp*, encoding the histone acetyltransferase CREB-binding protein (CBP). CBP is a transcriptional co-activator for CREB, and induction of CREB-dependent transcription plays a key role in long-term memory (LTM). In animal models of RTS, mutations of *cbp* impair LTM and late-phase long-term potentiation (LTP). As a step toward exploring plausible intervention strategies to rescue the deficits in LTP, we extended our previous model of LTP induction to describe histone acetylation and simulated LTP impairment due to *cbp* mutation. Plausible drug effects were simulated by model parameter changes, and many increased LTP. However no parameter variation consistent with a biochemical effect of a known drug class fully restored LTP. Thus we examined paired parameter variations consistent with effects of known drugs. A pair that simulated the effects of a phosphodiesterase inhibitor (slowing cAMP degradation) concurrent with a deacetylase inhibitor (prolonging histone acetylation) restored normal LTP. Importantly these paired parameter changes did not alter basal synaptic weight. A pair that simulated the effects of a phosphodiesterase inhibitor and an acetyltransferase activator was similarly effective. For both pairs strong additive synergism was present. The effect of the combination was greater than the summed effect of the separate parameter changes. These results suggest that promoting histone acetylation while simultaneously slowing the degradation of cAMP may constitute a promising strategy for restoring deficits in LTP that may be associated with learning deficits in RTS. More generally these results illustrate how the strategy of combining modeling and empirical studies may provide insights into the design of effective therapies for improving long-term synaptic plasticity and learning associated with cognitive disorders.

# INTRODUCTION

The identification of molecular lesions associated with various neurological disorders that adversely affect cognitive function is providing new opportunities for developing therapeutic interventions. The obvious first choice is to reverse the molecular lesion, or ameliorate its effects, with gene targeting techniques (Corti et al., 2012; Popiel et al., 2012; Rafi et al., 2012). Progress has been made, but this strategy has yet to provide a therapy for a human neurological disorder. A complementary approach is to pharmacologically manipulate an element in the biochemical pathway associated with the molecular lesion to compensate for loss of function (Park et al., 2014; Ehlinger et al., 2011; Guilding et al., 2007; McBride et al., 2005). A key challenge, however, is to identify the optimal site to target. In addition, it is increasingly clear from work in other fields that the targeting of multiple sites simultaneously may yield several advantages over targeting single sites. For example, current pharmacotherapies for cancer and infections often are combination therapies (Bijnsdorp et al., 2011). The majority of these therapies were developed by empirical, trial-and-error methods. However, there is a growing realization that a combination of empirical studies and modeling of intracellular signaling pathways can greatly aid the prediction of effective combinations of drugs (Boran and Iyengar, 2010; Severyn et al., 2011; Zhang et al., 2014). Combination therapies also may offer the advantages of synergism, which can allow lower dosages of the individual drugs to produce a desired effect (Barrera et al., 2005). Dose reductions due to synergism may also minimize undesirable off-target effects of the individual drugs (Zimmerman et al., 2007).

To examine the ways in which computational models and analyses of synergism may help to guide the development of therapies, we modeled aspects of the molecular network that underlies LTP, a neuronal correlate of memory. Several models have been developed to describe the dynamics of intracellular signaling pathways necessary for the induction of LTP (*e.g.* Bhalla and Iyengar, 1999; Hayer and Bhalla, 2005; Smolen et al., 2006, 2012). These models use empirical estimates of biochemical parameters such as cellular enzyme concentrations, Michaelis constants, and binding affinities. The convergence of multiple kinases and their downstream transcription factors to activate gene expression necessary for late LTP and the establishment of LTM, is likely to generate regions of synergism in these models. In the present study, we simulated effects of a molecular lesion associated with a cognitive disorder and LTP deficits, and attempted to predict drug combinations that could restore normal LTP and also exhibit synergism.

We selected as an exemplar Rubinstein-Taybi syndrome (RTS), a congenital disorder that is associated with cognitive and learning disability. RTS is associated with mutations in the gene encoding CREB binding protein (CBP), a histone acetyltransferase and an obligatory cofactor in the activation of transcription by cyclic AMP response element binding protein (CREB) (Barco, 2007; Graff and Mansuy, 2009; Park et al. 2014; Petrif et al., 1995; Roelfsema and Peters, 2007) (a small percentage of RTS is due to mutations in a related histone acetyltransferase, p300). In neurons, activation of protein kinase A (PKA) leads to CREB phosphorylation and consequent activation (Matsushita et al., 2001). Some forms of late LTP and LTM require activation of CREB (Kida, 2012; Peters et al., 2009; Pittenger et al., 2002) and co-activation of CBP (Korzus et al., 2004; Levenson et al., 2004). We explored whether our previous model describing LTP induction (Smolen et al., 2006), extended to include CBP and acetylation, could: a) simulate the impaired LTP seen in rodent models for RTS, b) suggest modulation of specific biochemical parameters as potential targets to rescue the deficit in LTP, c) identify pairs of parameters that are plausible drug targets and, when concurrently varied, rescue the deficit in LTP, and d) predict regions of synergism associated with concurrent adjustments to these parameter pairs. For RTS, two synergistic pharmacological manipulations were predicted to rescue LTP.

**METHODS**

**The extended model of LTP induction**

Our model was constructed to simulate induction of late, protein-synthesis dependent LTP at Schaffer collateral synapses in the CA1 region of the hippocampus. An overview of the model, and of stimulus inputs, follows. Ordinary differential equations describe the dynamics of kinase activities, gene expression, and synaptic weight. As schematized in Fig. 1, stimuli activate three signaling pathways. Increased [$Ca^{2+}$] activates CaM kinase II (CaMKII), Ras activation leads to activation of the ERK isoforms of MAPK, and increased [cAMP] activates PKA. These pathways are established as essential for LTP. PKA phosphorylation of an unspecified substrate is necessary to set a synaptic tag required for synaptic "capture" of proteins necessary for late LTP (Barco et al., 2002; Frey and Morris, 1997; Redondo and Morris, 2011). This PKA-sensitive synaptic tag is denoted Tag-2 (Fig. 1). Phosphorylation of another substrate Tag-1 by CaMKII is also necessary to set the tag (Fig. 1) as data suggests (Chen et al., 2001). Inhibition of ERK blocks LTP (English and Sweatt, 1997; Rosenblum et al., 2002), thus ERK phosphorylates another tag substrate, Tag-3. The value of a tag variable is set proportional to the product of the levels of Tag-1 – Tag-3. ERK is known to phosphorylate transcription factors (TFs), inducing genes important for LTP (Impey et al., 1998; Waltereit et al., 2001). In the model activated ERK phosphorylates a TF denoted TF-1. Activation of CREB is essential for at least some forms of late LTP. Inhibiting PKA attenuates CREB phosphorylation and LTP. Thus a second TF, TF-2, represents CREB and is phosphorylated by PKA.

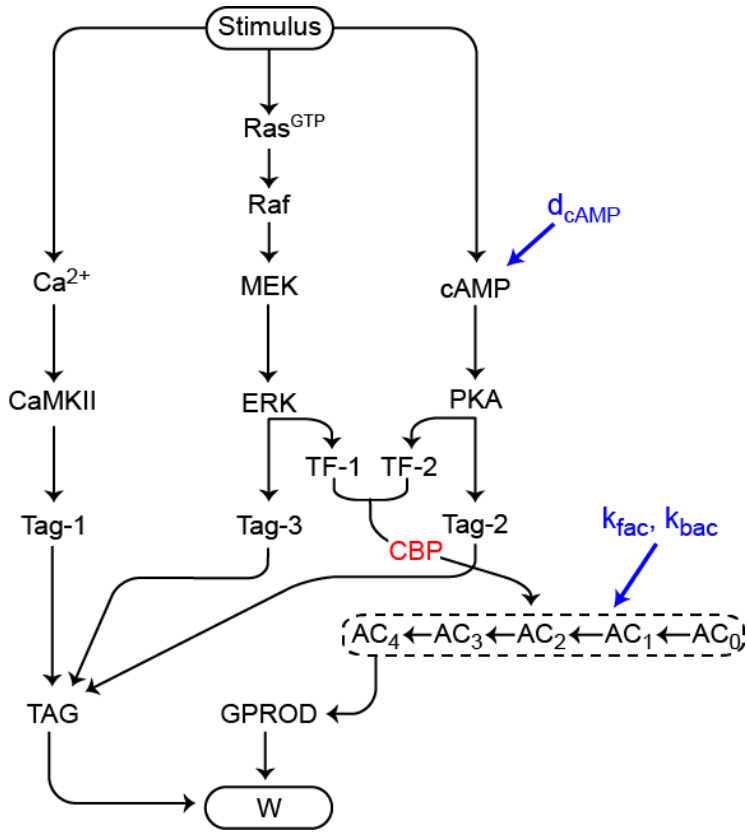

FIGURE 1 Model of key postsynaptic signaling pathways that underlie the induction of late LTP and of the mutation linked to RTS. Arrows indicate the sites of action of the key parameters used to model drug effects – $k_{fac}$ for acetyltransferase activators, $k_{bac}$ for deacetylase inhibitors, and $d_{cAMP}$ for PDE inhibitors.

We model histone acetylation as necessary for expression of a representative gene product necessary for LTP, denoted GPROD. Acetylation is included because the essential CREB cofactor CBP drives acetylation and gene induction and is deficient in RTS. In the model four acetylations are needed to induce GPROD synthesis. Both TF-1 and TF-2 need to be phosphorylated to induce these acetylations. The state with four acetylations is denoted AC4 (Fig. 1). The rate of GPROD expression is proportional to the amount of AC4. LTP is, then, modeled as an increase in a synaptic weight W. The rate of increase is simply taken as proportional to the product of the tag variable with GPROD.

In simulations LTP was induced by spaced tetani. Stimulus inputs were not modeled as differential equations, but rather as square-wave increases in synaptic $Ca^{2+}$, [cAMP], and Ras activity. Four 1-sec tetani at intervals of 5 min were simulated. This protocol was previously used with an RTS mouse model (Alarcon et al. 2004). Each tetanus induced an elevation of [cAMP] with duration $d_{cAMP}$. $d_{cAMP}$ is a key model parameter because increased $d_{cAMP}$ was used to simulate the effect of a phosphodiesterase inhibitor such as rolipram. Its control value was 1 min.

Further details of the model, and of stimulus parameters, are as follows. Equations and parameters are as in Smolen et al. (2006) except for two revisions discussed below. Separate synaptic and somatic ERK cascades are modeled. Equations and parameters describing these cascades are identical except that active somatic ERK can be imported into a nuclear compartment, with this import facilitated by PKA. Nuclear ERK phosphorylates TF-1. The original model assumed the second TF, TF-2, was phosphorylated by CaM kinase IV (CaMKIV). However, data shows that inhibiting nuclear PKA attenuates both CREB phosphorylation and LTP (Matsushita et al., 2001). Thus in the first revision TF-2, which represents CREB, was assumed to be phosphorylated by PKA, and CaMKIV and nuclear $[Ca^{2+}]$ were removed from the model. The phosphorylated fraction of TF-2 is denoted by a variable [TF2-P] ranging from 0 to 1, and governed by the following differential equation,

$$\frac{d\,[\text{TF2-P}]}{dt} = k_{phos} \text{PKA}_{ACT} \left(1 - [\text{TF2-P}]\right) - k_{deph}[\text{TF2-P}] \qquad 1)$$

The corresponding variable for TF-1, [TF1-P], similarly ranges from 0 to 1. TF-1 represents a TF other than CREB that is phosphorylated by ERK. Examples are Ets-1 or Ets-2, which when phosphorylated recruit CBP (Foulds et al., 2004). The combined action of TF-1 and TF-2 drives the production of a gene product assumed necessary for LTP, its concentration is denoted [GPROD].

The second revision was inclusion of histone acetylation by cooperation between CREB and CBP (Barrett et al., 2011). Multiple acetylations are believed to induce an "open" chromatin conformation allowing gene induction (Zentner and Henikoff, 2013). The model assumes four sequential acetylations, four differential equations and a conservation equation describe the kinetics of five histone states (0-4 acetylations). A forward acetylation rate constant $k_{fac}$ is proportional to the product of the concentrations of [TF1-P] and [TF2-P]. A first-order rate constant for deacetylation, $k_{bac}$, applies to all acetylated states. The equations describing histone acetylation are therefore as follows,

$$\begin{aligned}\frac{d\,(\text{AC1})}{dt} =\ & -k_{fac}(\text{AC1})\frac{[\text{TF1-P}]}{[\text{TF1-P}]+K_{A1}}\frac{[\text{TF2-P}]}{[\text{TF2-P}]+K_{A2}} + k_{bac}(\text{AC2}) \\ & + k_{fac}(\text{AC0})\frac{[\text{TF1-P}]}{[\text{TF1-P}]+K_{A1}}\frac{[\text{TF2-P}]}{[\text{TF2-P}]+K_{A2}} - k_{bac}(\text{AC1})\end{aligned} \qquad 2)$$

$$\frac{d(AC2)}{dt} = -k_{fac}(AC2)\frac{[TF1\text{-}P]}{[TF1\text{-}P]+K_{A1}}\frac{[TF2\text{-}P]}{[TF2\text{-}P]+K_{A2}} + k_{bac}(AC3)$$
$$+ k_{fac}(AC1)\frac{[TF1\text{-}P]}{[TF1\text{-}P]+K_{A1}}\frac{[TF2\text{-}P]}{[TF2\text{-}P]+K_{A2}} - k_{bac}(AC2) \quad 3)$$

$$\frac{d(AC3)}{dt} = -k_{fac}(AC3)\frac{[TF1\text{-}P]}{[TF1\text{-}P]+K_{A1}}\frac{[TF2\text{-}P]}{[TF2\text{-}P]+K_{A2}} + k_{bac}(AC4)$$
$$+ k_{fac}(AC2)\frac{[TF1\text{-}P]}{[TF1\text{-}P]+K_{A1}}\frac{[TF2\text{-}P]}{[TF2\text{-}P]+K_{A2}} - k_{bac}(AC3) \quad 4)$$

$$\frac{d(AC4)}{dt} = k_{fac}(AC3)\frac{[TF1\text{-}P]}{[TF1\text{-}P]+K_{A1}}\frac{[TF2\text{-}P]}{[TF2\text{-}P]+K_{A2}} - k_{bac}(AC4) \quad 5)$$

$$(AC0) = AC_{TOT} - (AC4) - (AC3) - (AC2) - (AC1) \quad 6)$$

The rate of GPROD expression is proportional to AC4. The differential equation describing GPROD expression is,

$$\frac{d[GPROD]}{dt} = k_{syn}(AC4) - k_{deg}[GPROD] + k_{synbas} \quad 7)$$

The rate of increase of the synaptic weight W is assumed proportional to the product of the tag variable with the gene product and is also limited by the availability, for synaptic incorporation, of another protein P (Smolen et al., 2006). The differential equations for W and P are as follows,

$$\frac{dW}{dt} = k_W(TAG)[GPROD]\frac{[P]}{[P]+K_P} - W/\tau_W \quad 8)$$

$$\frac{dP}{dt} = k_P(TAG)[GPROD]\frac{[P]}{[P]+K_P} + v_P - P/\tau_P \quad 9)$$

Empirically and in the model, tetanic stimuli elevate $[Ca^{2+}]$ and [cAMP] and activate the ERK cascade. Basal synaptic $[Ca^{2+}]$ was 0.04 μM. The effect of each tetanus (100 Hz for 1 sec) was simply modeled as concurrent square-wave increases in $[Ca^{2+}_{syn}]$, [cAMP], and Raf activation (Smolen et al., 2006). Each tetanus induced an increase of synaptic $Ca^{2+}$ to 1 μM, for 3 s. A similar duration is suggested by data (Pologruto et al., 2004). Each tetanus induced a square-wave elevation of [cAMP] with duration $d_{cAMP}$. Values for [cAMP] were 0.03 µM (basal) and 0.15 µM (elevated). Each tetanus also increased a rate constant $k_{f,Raf}$ for Raf phosphorylation / activation, for 1 min. $k_{f,Raf}$ was 0.0075 min$^{-1}$ (basal) and 0.16 min$^{-1}$ (elevated).

Standard model parameter values in the absence of simulated drug effects are given in Smolen et al. (2006), with the exception of the following parameters in Eqs. 1 - 9:

$k_{phos}$ = 0.12 µM$^{-1}$ min$^{-1}$, $k_{deph}$ = 0.03 min$^{-1}$, $k_{syn}$ = 1.0 µM min$^{-1}$, $k_{deg}$ = 0.01 min$^{-1}$,

$k_{synbas}$ = 0.0004 min$^{-1}$, $k_{fac}$ = 5.0 min$^{-1}$, $k_{bac}$ = 0.1 min$^{-1}$, $K_{A1}$ = $K_{A2}$ = 1.0, $AC_{TOT}$ = 1.0, $k_W$ = 4.0 min$^{-1}$,

$\tau_W$ = 100,000 min, $k_P$ = 15.0 µM$^{-1}$ min$^{-1}$, $v_P$ = 0.0006 µM$^{-1}$ min$^{-1}$.

The dynamics of histone acetylation and deacetylation following LTP induction have not been well characterized, and TF-1 is a generic transcription factor. Therefore, standard values for parameters in Eqs. 1 – 9 were chosen only according to the following criteria. GPROD induction was required to have a similar rate of increase, and duration, to that observed for induction of immediate-early genes associated with LTP, such as Arg3.1/Arc (Waltereit et al., 2001; Hevroni et al., 1998). The increase in W was required to develop over 2 h, similar to the time course observed for induction of chemical LTP (Yang et al., 2002). The amplitude of the W increase was required to be similar to that observed after three to four 1-s duration tetani (English and Sweatt, 1997; Woo et al, 2000). The very large time constant for the decay of W to baseline was chosen because of observations that late LTP can persist for months in vivo (Abraham et al., 2002).

The model of Smolen et al. (2006), with standard parameter values that simulate tetanic LTP, has been deposited in the online database ModelDB (accession # 149715).

**Numerical methods**

The simulations of Figs. 2 and 3 were integrated by two methods, forward Euler and fourth-order Runge-Kutta (Burden and Faires, 2005). No significant differences were observed in the results, therefore forward Euler was used for the remaining simulations. The time step was 40 ms. Prior to stimuli, model variables were equilibrated for at least two simulated days and the slowest variable, W, was set to an equilibrium basal value determined by the remaining variables. Programs are available upon request.

# RESULTS

## Simulations of LTP and of RTS-induced LTP deficits

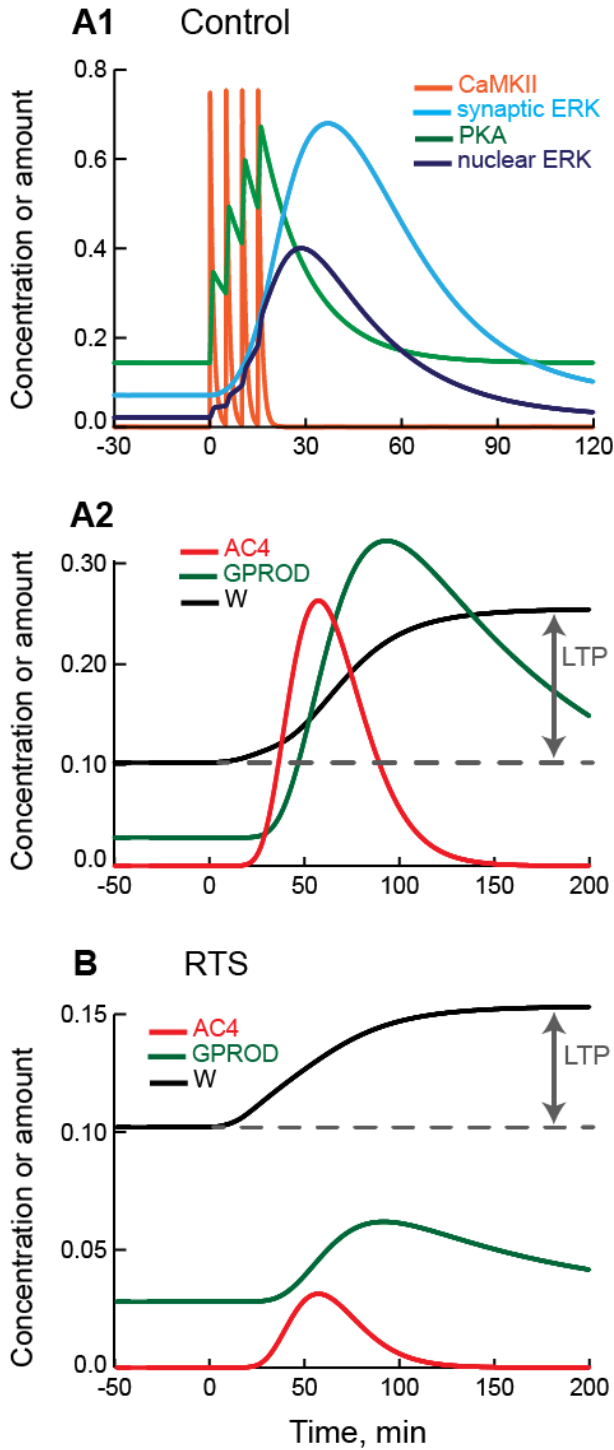

Figure 2 illustrates dynamics of some key variables of the model, with control parameter values, in response to the above LTP protocol. CaMKII responds rapidly to each brief $Ca^{2+}$ increase and deactivates in ~1 min, whereas synaptic ERK activity and PKA activity accumulate over tetani (Fig. 2A1). Nuclear import of ERK is driven by active PKA and after PKA deactivation nuclear ERK levels begin to drop. Thus nuclear ERK activity peaks before synaptic ERK activity. AC4, representing the transcription-competent chromatin state with four histone acetylations, peaks later than the kinase activities (Fig. 2A2). GPROD expression rises slowly.

FIGURE 2 Model dynamics during LTP induction. **A1** and **A2** illustrate the evolution of key variables during a simulation of normal LTP with all parameters at standard values. **B** illustrates a simulation of LTP when a parameter in the model was changed to simulate the CBP deficit associated with RTS. (**A1**) Time courses of the activities of CaMKII, synaptic ERK, nuclear ERK, and PKA in response to simulated tetani delivered at 0, 5, 10, and 15 min. (**A2**) Time courses of AC4, GPROD, and W. (**B**) Time courses of AC4, GPROD, and W for the RTS model. In all panels, time courses of all variables, except W, are vertically scaled for ease of visualization. Scaling factors for CaMKII, synaptic ERK, PKA, nuclear ERK, AC4, and GPROD are respectively 0.1, 7, 40, 15, 20, and 0.7. Also note changes in vertical scale for y axes of **A1**, **A2** and **B**.

The synaptic weight W increases for more than 2 h after stimulus, consistent with data illustrating that late LTP takes ~2 h to develop when induced by chemical stimuli that bypass early protein synthesis-independent LTP (Ying et al., 2002). The magnitude of simulated LTP was assessed 3 h after the last tetanus, as the per cent increase in W (Fig. 2A2 and 2B). Simulations using the control values for all parameters yielded LTP of 148% (Fig. 2A2).

In RTS, activity of the histone acetyltransferase CBP is reduced. In the model, the activity of CBP is represented by a forward rate constant for histone acetylation, $k_{fac}$ in Eqs. 2-5. To simulate RTS $k_{fac}$ was reduced from 5.0 min$^{-1}$ to 2.7 min$^{-1}$. The decrease in $k_{fac}$ only affects histone acetylation and downstream processes (GPROD synthesis and LTP). Figure 2B illustrates these dynamics (note difference in scale from Panel A). With the lower $k_{fac}$, much less AC4 accumulates, and GPROD synthesis is diminished. Therefore this case, denoted henceforth "RTS-basal", has reduced tetanic LTP – only 50%, compared to 148% in Fig. 2A2. This reduction by about two-thirds is similar to that observed in a rodent model of RTS, *cbp* heterozygous (*cbp*$^{+/-}$) mice, (Alarcon et al. 2004).

**Systematic parameter variations to rescue impaired LTP**

The predicted effect of drugs on LTP can be examined by varying the value(s) of model parameters that a candidate drug might change (e.g., a drug-induced reduction or increase in enzymatic activity). A successful candidate parameter change should restore LTP to near its normal value (148%) while basal W remains near its normal value. The latter criterion was imposed because extant literature does not appear to report abnormal basal weights (prior to LTP) in RTS.

Starting from the RTS-basal case (parameters at standard values except $k_{fac}$ = 2.7 min$^{-1}$), 15 parameters that represent plausible drug targets were varied. These parameters were: the durations of the stimulus-induced increases in Ca$^{2+}$, $k_{f,Raf}$, and cyclic AMP (cAMP); the CaMKII dissociation constant for Ca$^{2+}$/calmodulin denoted $K_{syn}$, the MEK and ERK activation Michaelis constants $K_{mkk}$ and $K_{mk}$, the Raf, MEK, and ERK inactivation rate constants $k_{braf}$, $k_{b,mapkk}$, and $k_{b,mapk}$; their activation rate constants $k_{f,mapkk}$, $k_{f,mapk}$, and $k_{f,raf}$ (basal); the PKA inactivation time constant $\tau_{PKA}$, the histone acetylation rate constant $k_{fac}$, and the opposing first-order rate constant for histone deacetylation in Eqs. 2-5, denoted $k_{bac}$. The duration of stimulus-induced cAMP elevation is denoted $d_{cAMP}$. Other model parameters were not varied because: a) they represent phosphorylation of unspecified targets, or synthesis and depletion of unspecified proteins, and thus cannot be targeted by specified drugs, or b) they represent overall synaptic weight dynamics and are not attributable to specific modeled biochemical processes. Also, parameters that regulate the basal level of Ca$^{2+}$ were not varied because changes in basal level of Ca$^{2+}$ would regulate too many processes to be a desirable drug target.

Each parameter was varied in turn by the same relative amounts, ±30%. This moderate amplitude of parameter change was chosen to represent plausible effects of moderate drug "doses". LTP was simulated for the RTS-basal case and for these 30 parameter variations. Normal LTP was also simulated (parameters at standard values, $k_{fac}$ = 5.0 min$^{-1}$). For each simulation, LTP (per cent increase of W from basal) was plotted *vs*. the basal weight (W prior to stimulus). Figure 3 illustrates the results. The red asterisk at (0.10, 148) denotes normal basal weight and LTP and the yellow square at (0.10, 50) denotes the simulated LTP deficit associated with RTS. We sought parameter variations that would restore LTP to near-normal values while maintaining basal synaptic weight near its normal value of 0.10.

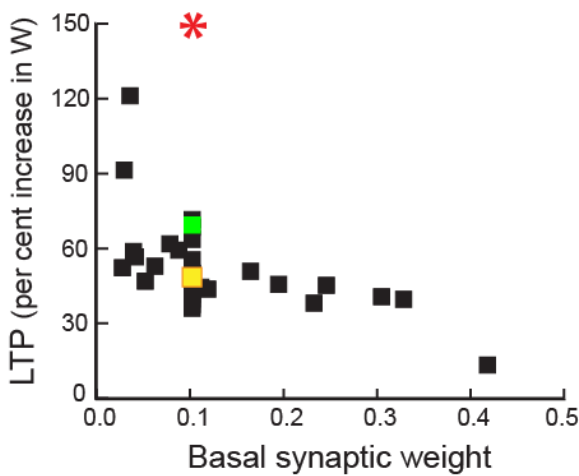

FIGURE 3 Variations of the magnitude of tetanic LTP with respect to changes in individual parameters of the model of Fig. 2. Starting from the RTS-basal simulation (yellow square), each model parameter that represents a plausible drug target is increased or decreased by 30%. The black squares and the green square illustrate the resulting LTP amplitudes *vs*. basal synaptic weights. Fifteen parameters are varied, yielding 30 values plotted as squares. The green square represents the effect of increasing $d_{cAMP}$ by 30% (the effect of decreasing $k_{bac}$ by 30% overlaps with this square) The asterisk at (0.10, 148) denotes the control simulation, with normal basal weight and LTP.

None of the single-parameter variations in Fig. 3 restored LTP to near-normal values. A 30% increase in the parameter $d_{cAMP}$, which denotes the duration of tetanus-induced [cAMP] elevation (from 1 to 1.3 min) increased LTP somewhat, to 69%, while leaving basal W at 0.102 (green square). This result plausibly corresponds to an effect of a PDE inhibitor, because inhibition of cAMP phosphodiesterase would prolong stimulus-induced cAMP elevation. Indeed, with *cbp*$^{+/-}$ mice, both LTP and learning have been improved by rolipram (Alarcon et al., 2004; Bourtchouladze et al., 2003), an inhibitor of cAMP phosphodiesterase 4 (PDE4) (Rose et al., 2005). By increasing cAMP levels, rolipram promotes PKA activation and CREB phosphorylation. A second PDE4 inhibitor, HT0712, also rescued learning (Bourtchouladze et al., 2003).

A 30% reduction in the rate constant for histone deacetylation, $k_{bac}$ in Eqs. 2-5, gave LTP of 68%. This result might correspond to the effect of a deacetylase inhibitor, acting to counter the effect of a reduced level of CBP. Indeed, in *cbp*$^{+/-}$ mice both LTP and learning are improved by SAHA, a histone deacetylase inhibitor (Alarcon et al., 2004). However, no ±30% change to a single parameter restored

LTP to >100%, except for increasing the dissociation constant for $Ca^{2+}$/calmodulin from CaM kinase II, $K_{syn}$. That increase yielded LTP of 121%, but at a cost of reducing the basal synaptic weight far below normal, to 0.036.

We also simulated larger increases in $d_{cAMP}$. A much larger, 100% increase elicited near-normal LTP (143%) while preserving basal W at 0.10 (not shown). However, such a large increase may correspond to a PDE inhibitor dose that would generate unacceptable side effects. With $cbp^{+/-}$ mice, rolipram did not completely restore normal LTP, although it yielded significant partial rescue (Alarcon et al., 2004). With our model, because single parameter changes did not give optimal results, we examined concurrent variations in pairs of parameters. Can LTP and basal synaptic weight both be restored by concurrent, moderate variations of two parameters?

Rather than exhaustively investigating each of the 105 distinct parameter pairs for the 15 parameters, we used existing knowledge of biochemical pathways to eliminate variations of parameters that *a priori* are likely to correspond to unacceptable off-target effects of drugs. We did not simulate the effects of increases in the activation rate constants for Raf, MEK, and ERK, or decreases in their inactivation rate constants. This constraint was imposed because excessive activation of the ERK signaling pathway is associated with tumorogenesis, and drug development has therefore focused on inhibitors of this pathway (Roberts and Der, 2007). Similarly, decreases in the Michaelis constants for activation of ERK or MEK by MEK or Raf were not simulated, because these parameter changes would also increase basal activation of the ERK pathway. Increases in the PKA inactivation time constant $\tau_{PKA}$ were also not included because no small-molecule, allosteric effector of PKA has been reported to increase this parameter. Alterations in the CaMKII dissociation constant $K_{syn}$, as noted above, greatly altered basal synaptic weight and were therefore not included. Applying these constraints left only pairs of four parameters to be examined. These parameters were the durations of the stimulus-induced increases in $Ca^{2+}$ and in cAMP, and the histone acetylation and deacetylation rate constants $k_{fac}$ and $k_{bac}$.

Simulations with the above constraints identified two parameter combinations that may represent candidate targets for rescuing deficits in LTP associated with RTS. In the first parameter pair, an increase in the duration of stimulus-induced cAMP elevation, $d_{cAMP}$, combined with a decrease in the histone deacetylation rate constant $k_{bac}$ restored LTP while preserving normal basal synaptic weight (0.10). With $d_{cAMP}$ increased by 50% and $k_{bac}$ decreased by 35%, LTP is 142%. These changes may respectively represent effects of a PDE inhibitor and a deacetylase inhibitor. In the second pair, a 50% increase in $d_{cAMP}$ in combination with a 37% increase in the histone acetylation rate constant $k_{fac}$ produced, LTP near normal (156%) and basal synaptic weight remained at 0.1. These changes may respectively represent effects of a PDE inhibitor and an acetyltransferase activator. Because these rescues appeared encouraging, we examined whether these pairs of parameters exhibit synergism. Qualitatively, synergism implies that drugs reinforce each other such that their effect in combination exceeds the prediction given by adding their separate effects (Bijnsdorp et al., 2011).

## Strong additive synergism occurs for both candidate parameter pairs.

We started with $k_{fac}$ = 2.7 min$^{-1}$ (the RTS-basal case), and simulated parameter-response (PR) curves for ($d_{cAMP}$, $k_{bac}$) (Fig. 4A). The response measure was percent LTP.

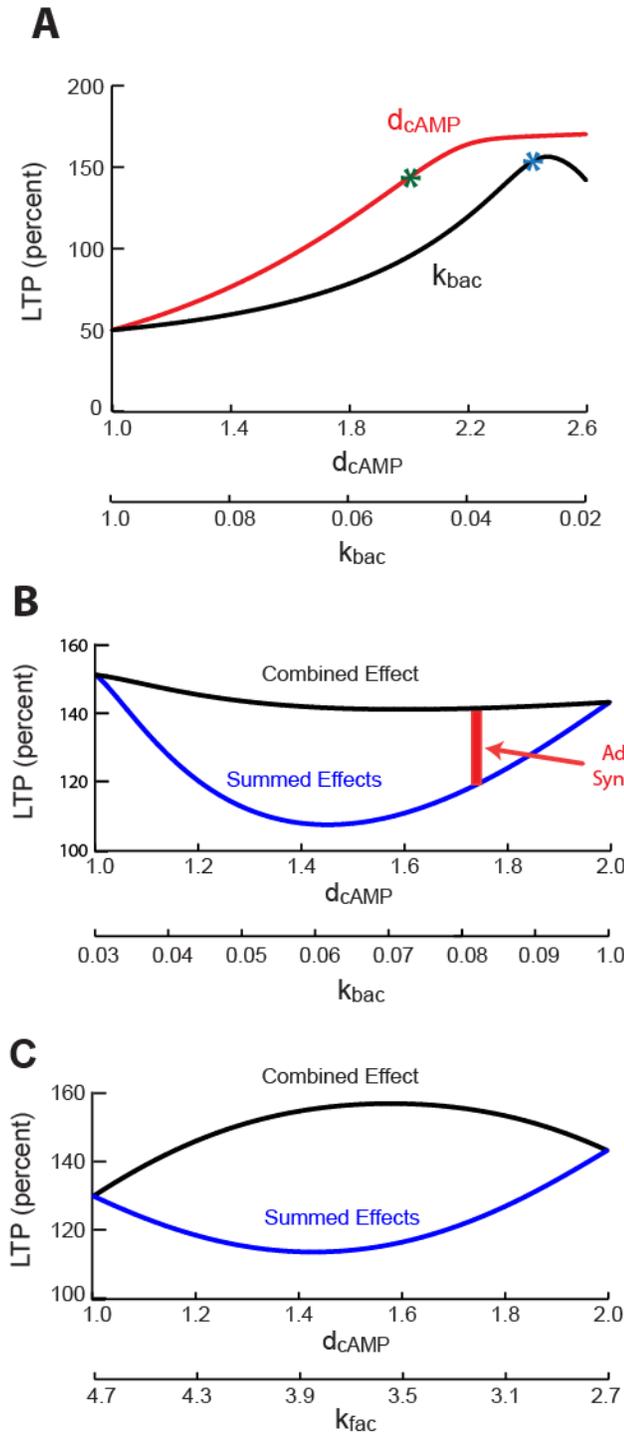

FIGURE 4 Simulated rescue of LTP deficits. (**A**) Parameter-response (PR) curves for the individual parameters, with maximal parameter changes chosen such that the magnitude of LTP is near normal (i.e., so that the simulated LTP defect is overcome). The green asterisk on the upper curve corresponds to a $d_{cAMP}$ increase to 2.0 min (used in **B**), the blue asterisk on the lower curve corresponds to a $k_{bac}$ decrease to 0.03 min$^{-1}$ (used in **B**). For each parameter, to construct its PR curve, LTP was simulated at 51 equally spaced parameter increments including the endpoints. (**B**) Additive synergism is seen over the entire parameter range when $d_{cAMP}$ is varied from 1.0 min to a maximum of 2.0 min, and $k_{bac}$ is concurrently varied from a minimum of 0.03 min$^{-1}$ to 1.0 min$^{-1}$. To construct these curves, LTP was simulated at 51 equally spaced parameter

increments including the endpoints. (**C**) Additive synergism is also seen when $d_{cAMP}$ is varied from 1.0 min to a maximum of 2.0 min and $k_{fac}$ is concurrently varied from a maximum of 4.7 min$^{-1}$ to its standard, RTS-basal value of 2.7 min$^{-1}$.

The PR curves of Fig. 4A are similar to drug dose-response curves. They delineate ranges of variation of the histone deacetylation rate constant ($k_{bac}$), and of the duration of stimulus-induced elevation of cAMP ($d_{cAMP}$), that give substantial, up to near-saturating, enhancement of the response. These are the ranges over which useful synergism might be expected to occur. Maximal parameter changes from control values, at the right endpoints of the curves, were chosen such that the magnitude of LTP was in the normal range (i.e., so the simulated RTS defect was overcome). These are large, not moderate, changes from control values. The $d_{cAMP}$ curve is fairly linear up to a saturated plateau. The $k_{bac}$ curve shows nonlinearity, acceleration to a peak followed by a decline. This nonlinearity occurs because $k_{bac}$ affects multiple sequential deacetylation reactions.

We chose an intuitive measure of synergism. Additive synergism occurs whenever, given fixed doses of drugs A and B, the response to A and B combined exceeds the sum of the responses to A alone and to B alone. For a model's prediction of additive synergism to be of therapeutic interest, the simulated synergism should be robust in that it should persist over a range of variation of model parameters. We implemented a novel, relatively simple way of visualizing whether robust additive synergism is present for a parameter pair. This method is illustrated in Figs. 4B-C. As the parameters are concurrently varied, two curves are constructed, denoted "combined effect" and "summed effect".

For each of these curves in Fig. 4B, the right endpoint corresponds to $d_{cAMP}$ increased above basal while $k_{bac}$ remained at basal (high "PDE inhibitor", zero "deacetylase inhibitor"), and the left endpoint corresponds to $k_{bac}$ decreased below basal while $d_{cAMP}$ was at basal (high "deacetylase inhibitor", zero "PDE inhibitor"). From the PR curves, these endpoint values of $d_{cAMP}$ and $k_{bac}$ were chosen such that the corresponding enhancements of LTP were substantial, to normal values. These endpoint values, of 2.0 min for $d_{cAMP}$ and 0.03 min$^{-1}$ for $k_{bac}$, correspond respectively to 144% LTP on the $d_{cAMP}$ PR curve in Fig. 4A (green asterisk) and to 153% LTP% on the $k_{bac}$ PR curve (blue asterisk). On the synergism curves (Fig. 4B), from left to right, $d_{cAMP}$ and $k_{bac}$ increase linearly. Thus with exception of the endpoints, all points are associated with a concurrent $d_{cAMP}$ increase above its standard value and $k_{bac}$ decrease below its standard value. For each concurrent increase and decrease, the corresponding point on the "combined effect" curve, directly above, provides the amplitude of the simulated LTP. The corresponding point on the "summed effect" curve is, in contrast, the LTP amplitude expected if the individual LTP enhancements due to the $d_{cAMP}$ increase and $k_{bac}$ decrease simply added. This LTP amplitude is calculated from two simulations in which each of these parameter changes are applied separately. The resulting LTP enhancements are added together and then added to the LTP obtained with all parameters at standard values used to simulate the LTP deficit. Additive synergism is present when the "combined effect" curve lies above the "summed effect" curve, which is the case in Figs. 4B (red vertical bar). Therefore, additive synergism is robust in the parameter range illustrated in Fig. 4B. Throughout this range, the basal synaptic weight remained within 15% of its control value of 0.10.

We also found synergism for the second parameter pair that rescued LTP, an increase in $d_{cAMP}$ combined with an increase in $k_{fac}$. For these curves, the right endpoint corresponds to $d_{cAMP}$ increased to a maximum while $k_{fac}$ remained at its RTS-basal value of 2.7 min$^{-1}$ (high "PDE inhibitor", zero "acetylase activator"), and the leftmost point to $k_{fac}$ increased to a maximum while $d_{cAMP}$ was at basal (high "acetylase activator", zero "PDE inhibitor"). Relatively large maxima were chosen for both $d_{cAMP}$ (2.0 min, on the saturated plateau of the PR curve in Fig. 4A) and $k_{fac}$ (4.7 min$^{-1}$), in order that the resulting combined effect curve simulated full restoration of LTP (from 50% in the RTS-basal case to ~140-160%). Strong additive synergism was simulated (Fig. 4C) and was robust through this parameter range. Throughout the range, basal synaptic weight remained within 18% of the control value of 0.10.

**DISCUSSION**

A model of some of the key biochemical pathways required for the induction of LTP could simulate impaired LTP seen in rodent models of RTS and suggest modulation of specific biochemical parameters as potential targets to rescue this LTP deficit. Moderate changes in single parameters failed to rescue LTP, but rescue was produced by moderate simultaneous changes in two pairs of parameters that are plausible targets of currently available drug types. Furthermore, the model proved useful in predicting regions of synergism illustrated with concurrent adjustments to these parameter pairs (Fig. 4). The model is a simplification of the complex processes underlying LTP and memory and of the effects of a CBP deficit. Nonetheless we believe the simulations of LTP rescue and of additive synergism in Fig. 4 suggest candidate drug combinations for rescue of LTP deficits that, based on rodent models, may correlate with cognitive deficits seen in RTS. One drug combination would use a PDE inhibitor and a deacetylase inhibitor (DAI), the second would replace DAI by an acetyltransferase activator.

PDE inhibitors and DAIs have been tested individually in rodent models of RTS (Alarcon et al., 2004; Bourtchouladze et al., 2003) and DAIs have been suggested as candidate therapeutics for several human disorders, including RTS (Kazantsev and Thompson, 2008). Rolipram has not proved suitable for human therapy due to gastrointestinal side effects (Scott et al., 1991), but new PDE4 inhibitors, such as HT0712 and oglemilast (Giembycz, 2008; MacDonald et al., 2007), are undergoing clinical trials. Several small-molecule activators of histone acetyltransferase activity have been reported. The compound CTPB activates p300 (Mantelingu et al., 2007), an acetyltransferase and transcriptional co-activator that has similar structure to CBP (Marmorstein and Trievel, 2009). However, to our knowledge, an acetyltransferase activator has yet to be tested in animal models of either RTS or other cognitive disorders. Also, no drug combinations appear to have yet been tested in an animal RTS model, although Alarcon et al. (2004) previously suggested that combining a drug similar to rolipram with a DAI might consistute a candidate therapy for RTS. Such tests could determine whether, as our model suggests, rescue of LTP and learning might be achieved with substantially lower doses of the individual drugs, possibly ameliorating side effects.

We believe that the use of models to identify concurrent parameter changes that rescue LTP deficits, or other phenotypes, can substantially increase the efficiency of subsequent empirical studies by prioritizing which drug combinations to test first, and at which dose ranges. In this study, simulations combined with inductive reasoning not only identified two promising parameter pairs out of >100 possible pairs, but also suggested, for both pairs, concurrent parameter variations for which additive synergism was maximized and LTP was simultaneously restored to near-normal values. In Fig. 4B (for $d_{cAMP}$ and $k_{bac}$) and Fig. 4C (for $d_{cAMP}$ and $k_{fac}$) the variations with maximal synergism are near $d_{cAMP} = 1.5$. We note that for Fig. 4C, the combined-effect curve is concave down with the center higher than the endpoints. This parameter pair therefore exhibits additional synergism, termed strong nonlinear blending synergism (Peterson and Novack, 2007). Subsequent empirical studies might initially focus on these pairs and develop empirical descriptions of dose-effect relationships that allow translation of parameter changes in this range to drug doses.

In addition to combination drug therapies, a related type of synergism that may prove useful in treating cognitive disorders could occur between drugs and optimized training protocols. One study (Zhang et al., 2011) illustrated that, for a simple form of learning, modeling of the dynamics of biochemical interactions in sensory neurons successfully predicted a set of interstimulus intervals (ISIs) between repeated stimuli that enhanced the formation of LTM. Experiments verified that the predicted stimulus spacing increased long-term synaptic facilitation (LTF, a correlate of LTM). A subsequent study (Liu et al., 2013) impaired LTF between sensory and motor neurons by knocking down CBP expression – an *in vitro* analogue of the molecular lesion responsible for RTS. Then, an extended variant of the model of Zhang et al. (2011) was used to predict a rescue protocol, consisting of a set of variable ISIs between repeated stimuli that would restore LTF to normal levels. Experiments verified that this protocol rescued LTF. These results suggest that in mammals, training protocols with variable intervals could be predicted, with the help of models, to enhance or restore learning. A logical extension of these results will be to use the model to predict the effects of combining drugs with improved training protocols. Additive synergism between drugs and training could be assessed within this paradigm.

A limitation of such combined strategies will be the incompleteness of models describing LTP induction combined with empirical uncertainties in parameter values. In model construction, parameter values are generally not all available from a single experimental preparation, or a single species of animal. Instead, data from several types of preparations (*e.g.* slice and cell culture) and animals (*e.g.* rodents and primates) need to be used (Bhalla and Iyengar, 1999; Hayer and Bhalla, 2005; Smolen et al., 2006; Smolen et al., 2012). Molecule copy numbers per neuron are also difficult to determine for enzymes or other macromolecules involved in LTP. Although these limitations are substantial, we believe that the potential benefits of this strategy that combines modeling and empirical studies are considerable.

## ACKNOWLEDGEMENTS

We thank Y. Zhang for comments on an earlier version of the manuscript. Supported by NIH grant R01 NS073974.